# Cosmology and Astrophysics without Dark Energy and Dark Matter

Shlomo Barak and Elia M Leibowitz
School of Physics & Astronomy, Tel Aviv University Tel Aviv, 69978, Israel
arxiv:astro-ph v1          Submitted 14 September 2009

## Abstract

We show that there is no need for the hypothetical Dark Energy (DE) and Dark Matter (DM) to explain phenomena attributed to them.

In contrast to the consensus of the last decade, we show that $\dot{a}$, the time derivative of the cosmological scale factor, is a constant ($\dot{a} = \text{const}$, $\ddot{a} = 0$). We derive H(z), the Hubble parameter, as a function of the redshift, z. Based on H(z), we derive a curve of the Distance Modulus versus log(z). This curve fits data from supernovae observations, without any free parameters. This fit is as good as that obtained by current cosmology, which needs the free parameters $\Omega_M$ and $\Omega_\Lambda$.

We obtain these results by using the hitherto un-noticed fact that the global gravitational energy density, $\in_g$, in our Hubble Sphere (HS) is equal to the Cosmological Microwave Background (CMB) energy density, $\in_{CMB}$.

We derive the dynamic and kinematic relations that govern the motions of celestial bodies in and around galaxies. This derivation does not require any gravitating matter beyond the observed baryonic matter. The theoretical Rotation Curves (RC), resulting from these relations, fit observed RCs. We obtain these results by examining the interplay between the local $\in_g$, around a galaxy and $\in_{CMB}$, which causes the inhomogeneous and anisotropic space expansion around a galaxy.

## 1. Introduction

### 1.1. DM was Suggested Based on Newtonian Physics

Based on Newtonian Physics, DM was suggested long ago by Oort (1932) and Zwicky (1933) to explain the seemingly non-Newtonian dynamics within the Milky Way galaxy and in clusters of galaxies. In the seventies, the discovery of flat RCs in and around galaxies (Rubin and Ford 1970; Ostriker, Peebles and Yahil 1974) added support for the idea of DM. However, DM has never been conclusively identified nor directly detected.

In addition to the assumption of the existence of DM there is a consensus that space in our universe expands homogeneously and isotropically, both globally and locally. This expansion is believed to also occur in the interior of individual atoms (Davis et al 2003, 2007). It is further understood that material bodies are held together mainly by gravitation and ElectroMagnetic (EM) forces, and are not affected by space expansion. However, according to General Relativity (GR), gravitation is the contraction (curving) of space around masses. Hence it is reasonable to consider the possibility that this contraction affects space around galaxies by causing, locally, the expansion to be inhomogenous and anisotropic. This creates a halo of inhomogeneous space in which the central acceleration, at a given distance from the centre of the galaxy and towards it, is larger than in the case of no expansion. This is proven by the application of an extended Gauss Theorem to the case of deformed space around a mass. This proof yields our Extended Newtonian Gravitational Law (ENGL) which enables us to dispel the need for DM.

The kinematic relation obtained is the Tully-Fisher relation, resulting in Rotation Curves (RC) that fit observed RCs in galaxies (Figures 2,3). The expansion of the universe is homogeneous and isotropic only when viewed globally, whereas in and around galaxies it is inhomogeneous, and hence responsible for phenomena attributed to DM. Milgrom (1983), in his MOND phenomenological model, was one of the first to suggest that there is no need for DM.

### 1.2 DE was suggested to explain the <u>supposed</u> changes in the rate of expansion of the universe

Current understanding of the universe is predicated on the validity of General Relativity (GR), and FRW metric, through the Friedmann equations.

Based on this understanding, DE was suggested to explain the <u>supposed</u> changes in the rate of expansion of the universe, and is also used, together with DM, to explain its Euclidian nature (flatness), (Riess et al 1998, Perlmutter et al 1999). However, DE has never been conclusively identified or directly detected.

It is our understanding that space contraction is expressed by $\in_g$ whereas space dilation in our universe is expressed by $\in_{CMB}$. Note that both $\in_g$ and $\in_{CMB}$ being energy densities are actually pressures. We show that the calculated present value for the global $\in_g$, in our HS, equals $\in_{CMB}$. Section 1.3 explains that this result is not accidental. We derive H(z), the Hubble parameter as a function of the redshift, z, by equating the expressions for $\in_g$ and $\in_{CMB}$ as functions of the scale factor, a. From H(z), we derive a curve of the Distance Modulus versus log(z). This curve fits all available data from supernovae observations (Figure 1). This fit is obtained without any free parameters, whereas current cosmology needs the free parameters $\Omega_M$ and $\Omega_\Lambda$.

Note that GR considers the energy of EM waves as having the same positive contribution to curving as that of matter.

In contrast to current thinking, we suggest that the energy of EM waves contributes to the negative curving of space (dilating it, as we explain in Section 2). This is the kind of curving currently attributed to DE.

This suggestion, on the face of it, seems to contradict GR since it can be wrongly understood as implying that a beam of light bends away from a mass rather than towards it. This understanding arises from the mistaken belief that photons are <u>independent</u> particles and, as such, a negative curvature contribution of the energy of photons would imply an anti-gravitational equivalent mass. However, as we explain below, our suggestion does <u>not</u> mean that light bends away from a mass – clearly light bends towards a mass, as experience shows.

The situation is clarified if photons are considered, as they should be, as wavepackets and not as independent particles. Their velocity is determined by the permittivity and permeability of space in their tracks, which are affected by the presence of a large mass. Hence, they bend towards the mass despite their individual negative contributions to the curvature of space, which is negligible.

### 1.3    The Idea

The issues of DE and DM do not stand alone but must be considered within the broader framework of Physics. This section sketches our considerations, in their broad context. Section 2 elaborates.

There is a consensus that space is foamy, deformable and vibrating. There is some agreement that the energy density, $\epsilon_{ZPF}$, of the Zero Point Fluctuations (ZPF) determines (mainly) space density (number of cells per unit of volume) and elasticity. This is possible if space behaves non-linearly (anharmonically). This is analogous to thermal expansion of a solid which is determined by its vibrational energy density. Adding energy density to that of the $\epsilon_{ZPF}$ in a zone of space reduces its density (dilation), whereas reducing the $\epsilon_{ZPF}$ in a limited zone of space creates a higher density (contraction). We consider space vibrations to be the EM waves, whereas in the current paradigm there is only a coupling between the two.

The Standard Model considers Elementary Particles (EP) as point-like and structureless, and the String Theory as strings. However, if EPs have finite size, and more than one dimension, ZPF vibrations with a wavelength larger than twice their linear dimension cannot vibrate inside them. Thus some of the inner energy density migrates outside in the process of their creation (Casimir effect). The specific structure of an EP is qualitatively irrelevant to this result.

The result of this emigrated energy is two-fold:

1. Space is contracted in and around an EP. This contraction is gravitation. The energy density of the contracted space around the EP is the local gravitational energy density, $\epsilon_g$. (Section 2.5 discusses the issue of $\epsilon_g$ in GR.)

2. The migrated energy, added to the ZPF outside the EP, is spread all over the HS. This energy, added to the ZPF energy, is radiation and hence can be detected. This radiation is mainly the CMB.

It is now clear why the $\epsilon_{ZPF}$ does not "gravitate" – it simply sets the standard space density.

Over time, the $\epsilon_{CMB}$ got its Black Body spectrum due to its interaction with matter. The measured value of $\epsilon_{CMB}$ is $4.17 \cdot 10^{-13}$ erg cm$^{-3}$ whereas, as we show, the calculated value of the global gravitational energy density $\epsilon_g$ in our universe is $\epsilon_g \sim 3.5 \cdot 10^{-13}$ erg cm$^{-3}$. Since both $\epsilon_g$ and $\epsilon_{CMB}$ depend on $a^{-4}$, their equality is retained over time. Hence, equating the expressions for $\in_{CMB}$ and $\in_g$ as functions of the scale factor, a, enables us to derive our H(z) which, as was already mentioned, is supported by its fit to observations.

The equality of the present values of $\in_g$ to $\in_{CMB}$, and the fit of our H(z) to observations, support our idea.

Since $\in_g$ expresses a contracting pressure and $\in_{CMB}$ a dilating pressure, it is reasonable to assume that space expansion can take place only where and when $\in_{CMB} > \in_g$.

Globally, $\in_g$ is close to but less than $\in_{CMB}$ as explained below. However, locally, close to the center of a galaxy, as an example, $\in_g$ can be larger than $\in_{CMB}$. Hence it is understood that the expansion around the core of a galaxy in inhomogeneous and anisotropic. This is the basis for our derivation of the dynamic and kinematic relations for a celestial body moving around the core of a galaxy.

This result further strengthens our confidence in our idea.

The next section summarizes the main features of our idea relevant to this paper.

**1.4 Our Main Assumptions**

Our main assumptions are:

1. Space is three dimensional (3-D), foamy and deformable. The density and elasticity (the permittivity and permeability) of space are determined by its Zero Point Fluctuations (ZPF) and are affected by the presence of matter and radiation. This idea is not new.

2. The presence of matter causes space contraction (positive curving) around it – a higher density and a lower tension – larger permittivity and permeability. The gravitational energy density $\in_g$ expresses this contraction. This is the essence of GR.

3. The presence of radiation causes space dilation – a lower density and a higher tension. The EM energy density, $\in_{EM}$ expresses this dilation. This is a new idea, in distinct contrast to current understanding.

4. Space expands only where and when $\in_{EM} > \in_g$. In our discussion the relevant $\in_{EM}$ is $\in_{CMB}$. Space retains its standard density and tension due to the ZPF where and when $\in_{CMB} = \in_g$. This is the result of the properties attributed above to $\in_g$ and $\in_{EM}$.

GR does not take into account the interplay between $\in_g$ and $\in_{CMB}$ throughout the evolution of the universe, and hence is not useful in solving the issues of dark energy and dark matter. <u>This in no way implies that GR is not valid</u>. However, Assumption 3 requires a modification of the way that EM is incorporated in GR.

In this paper we validate our assumptions by the fit of our theoretical results with observed data.

## 2. Basic Concepts

This section discusses our assumptions.

### 2.1. Space is foamy

The consensus that space is foamy, and hence cellular, rests on the meaning of expansion, and the requirement that its vibrations have a finite energy density. By "its vibration" we mean the EM waves - this is not crucial to our discussion and appears simply as a remark. The cut-off wavelength of the ZPF, which determines its energy density, is the smallest linear dimension of a space cell. Whether this linear dimension is Planck's length, or not, is not relevant to our discussion.

It is interesting to note that B. Riemann, quoted by Chandrasekhar in Nature (1990), was of the opinion that space is foamy.

### 2.2. Space is three-dimensional and deformable

Our assumption that space is 3D means that the universe is <u>not</u> a curved 3D manifold in a hyperspace with an additional spatial dimension. For a globally flat universe the issue of an additional spatial dimension is not relevant. The terms "deformed" and "curved" are used for a 3D elastic space and a 3D-manifold, respectively. Note that Riemannian geometry is the geometry of both curved manifolds and deformed spaces. This is explained by A. Einstein (1921) and R. Feynman (1963).

The deformation of space is the change in size of its cells. Positive or negative deformation, around a point in space, means that the space cells grow or shrink, respectively, from this point outwards. For a positively curved manifold, the ratio of the circumference of a circle to the radius is less than $2\pi$, as measured by a yardstick of fixed length. For a deformed 3D-space, with a positive deformation, around a point the above ratio is also less than $2\pi$, as measured by a flexible yardstick such as the linear dimension of a space cell. Note that for a <u>deformed space</u> there is no meaning to global deformation, <u>deformation is a local attribute</u>. The surface of a sphere with radius R is a 2D manifold with a global curvature 1/R. However, a global homogeneous deformation for a deformable 2D planar sheet can only have the value zero i.e., the sheet is Euclidian. Here, deformation around a point is expressed by a scale factor $a(r,t)$ that depends on both time and the vector, r, from the point. Space density is reciprocal to $a(r,t)^3$.

### 2.3. Gravitation is the contraction of space due to the presence of a mass

GR shows that a mass curves space around it. This curving is the <u>contraction</u> of space around the mass. Cells close to the mass are smaller than those at a distance, and hence the elastic <u>positive</u> deformation of space. Length, close to a mass, is smaller, and the "running of time" is slower, than at a distance.

Gravitation is the elastic deformation of space, remove the mass and the deformation is gone. In Section 4 we discuss the issue of elastic versus non-elastic 3D space deformation and show that deformation due to space expansion is non-elastic.

Space contraction by a mass can be represented by a gravitational scale factor $a_g \equiv \ell/\ell_0$, where $\ell_0$ is a distance in space between two stationary points far from masses, and $\ell$ is the same distance contracted by the introduction of a mass. In GR, $\ell/\ell_0 = \exp(\varphi/c^2)$, where φ is the gravitational potential. Thus $a_g$ at the surface of the sun, or at the edge of our galaxy, is approximately $1 - 10^{-6}$ whereas in the last 11 BY the scale factor used in cosmology changed, due to expansion, from 0.25 to 1.

This difference, orders of magnitude, in space deformation is related to the elastic versus non-elastic behavior of space (see Section 4). This large deformation creates a need for an extended Newtonian gravitational law for galaxies, around which space expansion is inhomogeneous and anisotropic.

Note that space expansion of the universe is the enlargement of its space cells. The number of space cells in the universe is thus considered conserved.

**2.4. Space density is determined by its EM energy density, mainly by its ZPF**

Sakharov (1968) and Misner, Thorne and Wheeler (1970) suggested that the violent environment at the Planck scale determines the elasticity of space. Puthoff (1989) suggested that it is the ZPF that determines the elasticity of space. This implies that the ZPF also determines the size of a space cell. Note that in current thinking there is only coupling between space vibrations and the EM waves, as the bending of a beam of light close to a star indicates.

The success of Quantum Electro-Dynamics (QED) in predicting non-linear phenomena is an indication of the non-linearity of EM. This suggests that Maxwell's equations, being linear, are merely an approximation.

The scattering of light by light, without the presence of matter in the space occupied by the interacting beams, which results in the production of pairs of electrons and positrons (D. L. Burk et al, 1997 and G. Brodin et al, 2002) is an experimental indication of the non-linearity of space.

The above suggests that, as in solids, it is the anharmonicity (non-linearity) of space that enables vibrations to determine the size of its cells.

**2.5. Gravitational Energy (GE) is energy deducted from the energy of space without matter**

GE has a negative sign since it expresses the work done by the field bringing together material particles. This negative sign can be understood if we refer to the energy density of space, due to the ZPF, as formally set to be zero. The introduction of a mass contracts space around it. This elastic deformation is accompanied by a Displacement Vector field **u**. Gravitation is space deformation, as GR states, hence the gravitational field **E**$_g$ is proportional to **u**. This contractual deformation results in smaller space cells around the mass (positive curvature) with less tension. In the theory of elasticity this is expressed by the energy density $\in \propto \mathbf{u}^2$ (Landau and Lifshitz, 1986). In the theory of gravitation, GE density, which is the result of aggregation of material particles, is expressed by $\in_g = -E_g^2/8\pi G$ (J. Keller, 2006). In this paper, we relate only to the absolute value of $\in_g$.

GE, as we understand it, is the space elastic energy, due to the presence of mass, deducted from the energy of space, formally set to zero. Here we consider $\epsilon_g$ to be the inward contractual pressure of space. Section 2.4 considers the $\epsilon_{CMB}$ to be the outward dilational pressure of space. Hence, <u>by their nature</u>, $\epsilon_g$ and $\epsilon_{CMB}$ can be compared.

Note that in GR there is no well-defined gravitational field energy (L. D. Landau and E. M. Lifshitz 1962). Standard methods to obtain the energy-momentum tensor yield a non-unique pseudo-tensor. It is also asserted, based on the equivalence principle that the gravitational energy cannot be localized (C. W. Misner et al, 1970). However, recently, M. J. Dupré (2009) presented a fully covariant energy momentum stress tensor of the gravitational field which supports our understanding of $\epsilon_g$.

## 2.6. The cosmological principle (CP) implies different geometries of the universe for different types of space

For a 3D space manifold, curved in a hyperspace with an extra spatial dimension, CP implies a uniform global curvature. In this case the universe is finite but with no boundary.

For a 3D space CP implies flatness. This is the result of deformation being a local attribute only. In other words the only global curvature possible is zero. In this case the universe is either infinite, or finite with a boundary. Close to a boundary CP can not hold true.

## 3. Dark Energy

### 3.1. The inter-galactic gravitational energy density equals the CMB energy density

Gravitational energy is confined in each and every Hubble Sphere (HS) since gravitational contraction moves at the speed of light relative to the mass that is generating the gravitational field. Every point in the universe is both the center and the edge of some supposedly identical HSs. Therefore, the global $\epsilon_g$, far from masses, must be the same.

$$\epsilon_g = \frac{1}{8\pi G} E_g^2 = \frac{1}{8\pi G}\left(\frac{GM}{R_{HS}^2}\right)^2 \quad \text{hence:}$$

(1) $$\epsilon_g = \frac{2\pi G}{9} m^2 R_{HS}^2 = \frac{2\pi G}{9} m^2 \frac{c^2}{H^2}$$

M - HS mass, m – baryonic mass density of the universe, $R_{HS}$ – radius of an HS.

The present calculated value of the global $\epsilon_g$ taking m = $2 \cdot 10^{-31}$ gm cm$^{-3}$ based on Big Bang Nucleosynthesis (BBN) (Rindler, 2004) and the Hubble constant, $H_0 = 2.3 \cdot 10^{-18}$ s$^{-1}$, is:
$$\epsilon_g = 3.5 \cdot 10^{-13} \text{ erg cm}^{-3}$$
whereas the present measured value for the $\epsilon_{CMB}$ is:
$$\epsilon_{CMB} = 4.17 \cdot 10^{-13} \text{ erg cm}^{-3}$$

Both $\epsilon_g$ and $\epsilon_{CMB}$ depend on a$^{-4}$, hence their equality is retained over time and we can equate their expressions as functions of a.

The difference between the results for $\epsilon_g$ and $\epsilon_{CMB}$ can be explained by uncertainties in the observed values of m and H and the fact that $\epsilon_g$ is concentrated around masses, whereas $\epsilon_{CMB}$ is distributed homogeneously.

In the inter-galactic space, it seems that $\epsilon_{CMB} \gtrsim \epsilon_g$. It is not clear to us if this explains the expansion of the universe. In and around galaxies $\epsilon_g > \epsilon_{CMB}$ and hence expansion is inhibited in these regions. Section 4 shows that this inhomogeneous expansion results in the flattening of Rotation Curves (RC).

Inhabitants of a 3D universe can only make observations related to internal deformations. However, such deformations, on a global scale, do not appear since $\epsilon_g \sim \epsilon_{CMB}$. The result is the <u>Euclidian nature of the universe</u>, and hence the validity of CP, in general, or far from the boundary, if there is one.

In no way is a <u>critical mass density</u> or the <u>idea of inflation</u>, involved in our considerations and calculations.

**3.2. We derive the dependence (evolution) of the Hubble parameter, H, on the scale factor, a, by equating $\epsilon_{CMB}$ to the gravitational energy density $\epsilon_g$ in the universe**

From equation (1):

$$\epsilon_g = \frac{2\pi G}{9} m^2 \frac{c^2}{H^2} = \epsilon_{CMB}$$

Values as of today are designated by the index 0. Note that the scale factor, a, as of today is chosen as 1 and hence in the past was less than 1. The above equation gives:

(2) $\quad H^2 = \dfrac{2\pi G c^2 m^2}{9\, \epsilon_{CMB}}$

Substituting $m = m_0 \cdot a^{-3}$ and the known relation, $\epsilon_{CMB} = \epsilon_{CMB_0} \cdot a^{-4}$ (Frieman, Turner, Huterer 2008) in Equation (2) gives:

$$H^2 = \frac{2\pi G}{9} \frac{m_0^2 \cdot c^2}{\epsilon_{CMB_0}} \cdot a^{-2} \qquad \text{hence:}$$

$$H = m_0 \cdot c \sqrt{\frac{2\pi G}{9\, \epsilon_{CMB_0}}} \cdot a^{-1} \qquad \text{and thus:}$$

(3) $\quad H = H_0\, a^{-1} \qquad$ where:

(4) $\quad H_0 = m_0 \cdot c \sqrt{\dfrac{2\pi G}{9\, \epsilon_{CMB_0}}}$

Checking the result for $H_0$, using the present measured value for $\epsilon_{CMB}$, gives $H_0 = 2 \cdot 10^{-18}$ s$^{-1}$.

The recently measured value for $H_0$ is:
$\quad H_0 = (72 \pm 8)$ km s$^{-1}$/Mpc $= (2.3 \pm 0.26) \cdot 10^{-18}$ s$^{-1}$ (Freedman et al, 2000).

### 3.3. We derive the Hubble Parameter, H as a function of time, t, and show that $\dot{a} = \text{const}$ ($\ddot{a} = 0$)

This result complies with observation (Section 3.5). $H \stackrel{\text{def}}{=} \dot{a}/a$, but, as we have shown:
$H = H_0/a$, hence, $\dot{a} = \text{const}$ ($\ddot{a} = 0$).

The value of the constant $\dot{a}$ is $2.3 \cdot 10^{-18}$ sec$^{-1}$, since $H_0 = \dot{a}/a_0$, where $a_0 = 1$ today. Integrating both sides of $da = H_0 dt$, (since $a_0 = 1$), and designating BB - Big Bang, 0 - now, $t_{BB} = 0$, gives:

$$a - a_{BB} = H_0 t \qquad \text{but:} \quad a = \frac{H_0}{H} \quad \text{hence:}$$

(5) $\quad t = \dfrac{1}{H} - \dfrac{1}{H_{BB}} \qquad$ Now, at $t = t_0$, $H = H_0$ hence:

(6) $\quad t_0 = \dfrac{1}{H_0} - \dfrac{1}{H_{BB}} \qquad$ Since $H_{BB} \gg H_0$

$\quad t_0 \sim \dfrac{1}{H_0} = 13.7 \text{ BY} \quad$ the age of the universe

(7) $\quad H(t) = \dfrac{1}{t + 1/H_{BB}}$

In this derivation $H_{BB}$ is merely the result of the mathematics. In this paper, we do not relate to the issue of the BB.

The distance between any two galaxies grows with a, but H falls with a. We thus conclude that any two galaxies recede from each other at all times at a constant velocity $v = r \cdot H$.

### 3.4. Cosmological redshift due to space expansion is $z = e^{\frac{v_r}{c}} - 1$

The basis for our discussion is:

- A yardstick and a clock, far from masses are <u>not</u> affected by their location in space or by time.

- The distance between two points in an expanding universe, as measured by a fixed yardstick, is $d = \ell a$.
  a is the scale factor with its present chosen value $a = 1$ hence in the past $a < 1$.
  $\ell$ is defined as the ratio $d/a$. Alternatively, $d = na$ where n is the number of space cells of equal linear dimensions, and a is the linear dimension of a cell.

- Light velocity, c, is a constant of nature, affected only by the presence of mass.

- The distance between a photon, or crest of a wave, and its emitter is: $d(t) = \ell(t)a(t)$ and its velocity relative to the emitter (located in galaxy A) is:
  (8) $\quad v_p = \dot{d}(t) = \dot{\ell}(t) \cdot a(t) + \ell \cdot \dot{a}(t)$
  The first term on the right-hand side is the light velocity:
  (9) $\quad c = \dot{\ell}a$
  whereas the second term is the recessional velocity of the place at which the photon is momentary "present".

$v_r = \ell \dot{a}(t)$     therefore:

$v_p = c + v_r$     In this discussion, Special Relativity is not relevant.

We have shown that $\dot{a} = const$ hence:

(10)   $a(t) = a(t_z) + \dot{a}(t - t_z) = a_z + \dot{a}(t - t_z)$     for any time $t \geq t_z$.

In our discussion we use $t_z$ as the time of emission of our photon from galaxy A.

From $c = \dot{\ell} a$ we get:

$$\frac{d\ell}{dt} = \frac{c}{a} = \frac{c}{a_z + \dot{a}(t - t_z)}$$

$$\frac{1}{c} d\ell = \frac{dt}{a_z + \dot{a}(t - t_z)} \quad \text{and by integration:}$$

$$\frac{1}{c}(\ell - \ell_z) = \frac{1}{\dot{a}} \{\ln[a_z + \dot{a}(t - t_z)] - \ln a_z\} \quad \text{but } \ell_z = 0, \text{ hence:}$$

$$\frac{\ell}{c} = \frac{1}{\dot{a}} \cdot \ln\left[1 + \frac{\dot{a}}{a_z}(t - t_z)\right] \quad \text{but } \frac{\dot{a}}{a_z} = H(t_z) = H_z, \text{ hence:}$$

$$t - t_z = \frac{1}{H_z}\left[\exp\left(\frac{\dot{a}\ell}{c}\right) - 1\right]$$

Let $t_0$ be the cosmic time of arrival of a wavecrest to the observer in galaxy B.

$$t - t_z = \frac{1}{H_z}\left[\exp\left(\frac{\dot{a}\ell(t_0)}{c}\right) - 1\right] \quad \text{but } \ell(t_0) = \ell_0 = d_0 \text{ is the present distance}$$

between galaxies A and B. Hence $\dot{a}\ell_0$ is their recessional velocity $v_r$, and thus the Look Back Time is:

(11)   $(t - t_z) = \dfrac{1}{H_z}\left[e^{\frac{v_r}{c}} - 1\right]$

The cosmological redshift is the result of successive crests arriving at the observer with a longer arrival time difference $\Delta t'$ than their time difference $\Delta t$ at emission.

$\lambda_{em} = c \cdot \Delta t$    whereas    $\lambda_{obs} = c \cdot \Delta t'$.

$\Delta t' = t_2 - t_1$    $\Delta t = t_{z_2} - t_{z_1}$

$$\Delta t' = [(t_2 - t_{z_2}) - (t_1 - t_{z_1})] + \Delta t$$

Equation (11) gives:

(12)   $\Delta t' = \left[\dfrac{1}{H(t_2)} - \dfrac{1}{H(t_1)}\right] \cdot \left[e^{\frac{v}{c}} - 1\right] + \Delta t$     and since:

(13)   $H(t) = \dfrac{\dot{a}}{a} = \dfrac{\dot{a}}{a_z + \dot{a}(t - t_z)} = \dfrac{\dot{a}/a_z}{1 + \dot{a}/a_z(t - t_z)} = \dfrac{H}{1 + H_z(t - t_z)}$

our z as a function of the recessional velocity is:

(14) $\quad z = \dfrac{\lambda_{obs}}{\lambda_{em}} - 1 = e^{\frac{v_r}{c}} - 1$

For $v_r = c$ we get $z = 1.718$.

GR, using $\Omega_M = 0.3$ and $\Omega_\Lambda = 0.7$, gives for $v_r = c$, the value $z \geq 1.5$. See Figure 2 in Davis and Lineweaver (2003).

We derive the known relation, $a = 1/(1+z)$, from equation (9).

For dt the time difference between successive crests:

$$c = \dfrac{d\ell}{dt} \cdot a \qquad \text{gives:}$$

$$\lambda_{obs} = c \cdot dt_{obs} = d\ell \cdot a_{obs}$$

$$\lambda_{em} = c \cdot dt_{em} = d\ell \cdot a_{em} \qquad \text{hence:}$$

$$\dfrac{\lambda_{obs}}{\lambda_{em}} = \dfrac{a_{obs}}{a_{em}} \qquad \text{For } a_{obs} = 1 \text{ and } a_{em} \equiv a:$$

$$\dfrac{\lambda_{obs}}{\lambda_{em}} = \dfrac{1}{a}$$

$$z = \dfrac{\lambda_{obs}}{\lambda_{em}} - 1 = \dfrac{1}{a} - 1 \qquad \text{which gives:}$$

$$a = \dfrac{1}{1+z}$$

This is the result of differences in the arrival times of successive wave crests.

This result is valid only for the dynamic case of an expanding space between emitter and absorber.

### 3.5. The relations $H = H_0 a^{-1}$ and $a = 1/(1+z)$ give $H(z) = H_0(1 + z)$. We show that this result is confirmed by observations

In this section we use the conventional notation $H(z) = H_0 h(z)$. In <u>our</u> theory:

(15) $\quad h(z) = 1 + z$

Whereas the <u>known</u> equation with the two dependent free parameters, $\Omega_M$ and $\Omega_\Lambda$, for <u>flat space</u> where $\Omega_M + \Omega_\Lambda = 1$ is:

(16) $\quad h(z) = \left[ (1+z)^2 (1 + \Omega_M z) - z(2+z) \Omega_\Lambda \right]^{\frac{1}{2}}$

(Perlmutter, 1997).

Our h(z) yields a different Luminosity Distance (LD), $d_L$, from that derived from equation (16). LD is defined by the ratio of the luminosity, L, of a supernova, to its measured flux, F:

(17) $\quad d_L^2 \equiv \dfrac{L}{4\pi F}$

From the known relation:

(18) $\quad d_L = (1+z)\dfrac{c}{H_0} \cdot \int_0^z \dfrac{dz'}{h(z')} \quad$ using <u>our</u> h(z), we get:

(19) $\quad d_L = (1+z)\dfrac{c}{H_0} \cdot \int_0^z \dfrac{dz'}{1+z'} = (1+z)\dfrac{c}{H_0}\ln(1+z)$

Note that while the LD derived in FRW cosmology is a function with two dependent free parameters, $\Omega_M + \Omega_\Lambda = 1$ (flat universe) that can be, and were in fact adjusted to various data sets, in the last decade, our LD is obtained directly from theory, with <u>no free parameters</u>.

Figure 1 shows the distance modulus versus log(z), where z is the cosmological redshift.

**Frame (a)** displays the measured data in 307 SN Ia as compiled and presented in the website of the SNP Union http://supernova.lbl.gov/Union, Kowalski et al (2008).
The three curves are, from bottom up:

(i) (Green) derived from the FRW cosmology for a flat and matter-dominated universe with $(\Omega_M,\Omega_\Lambda) = (1.0, 0.0)$.

(ii) (Red) derived by our cosmology with no acceleration.

(iii) (Black) derived from the FRW cosmology for an accelerating flat universe with $(\Omega_M,\Omega_\Lambda) = (0.3, 0.7)$.

**Frame (b)** zoom on the upper-right corner of Frame (a), shows the fit of the three curves to the data.

**Frame (c)** extends the theoretical curves to $z = 100$. The family of thin lines are FRW lines for 11 pairs of the flat-universe parameters starting from $\Omega_M = 0.10$, $\Omega_\Lambda = 1 - \Omega_M$ in steps of 0.1. The three heavy lines are the same as in Frames (a) and (b). The highest thin line is that of an FRW $(\Omega_M,\Omega_\Lambda) = (0.0, 1.0)$ which represent a dark energy-dominated universe.

**Frame (d)** is a zoom on the upper-right corner of Frame (c). It shows the theoretical curves for large z values.

Note the cross-over between our curve (Red) and the curve (Black) of the currently accepted Λ CMB cosmology, which occurs around log(z) = 0.76, z = 5.76. Thus for higher z values our cosmology predicts a distance modulus as a function of log(z) curve that is distinctly different from that predicted by Λ CMB cosmology for a flat universe. Future measurements of high z standard candles should distinguish between the two cosmologies.

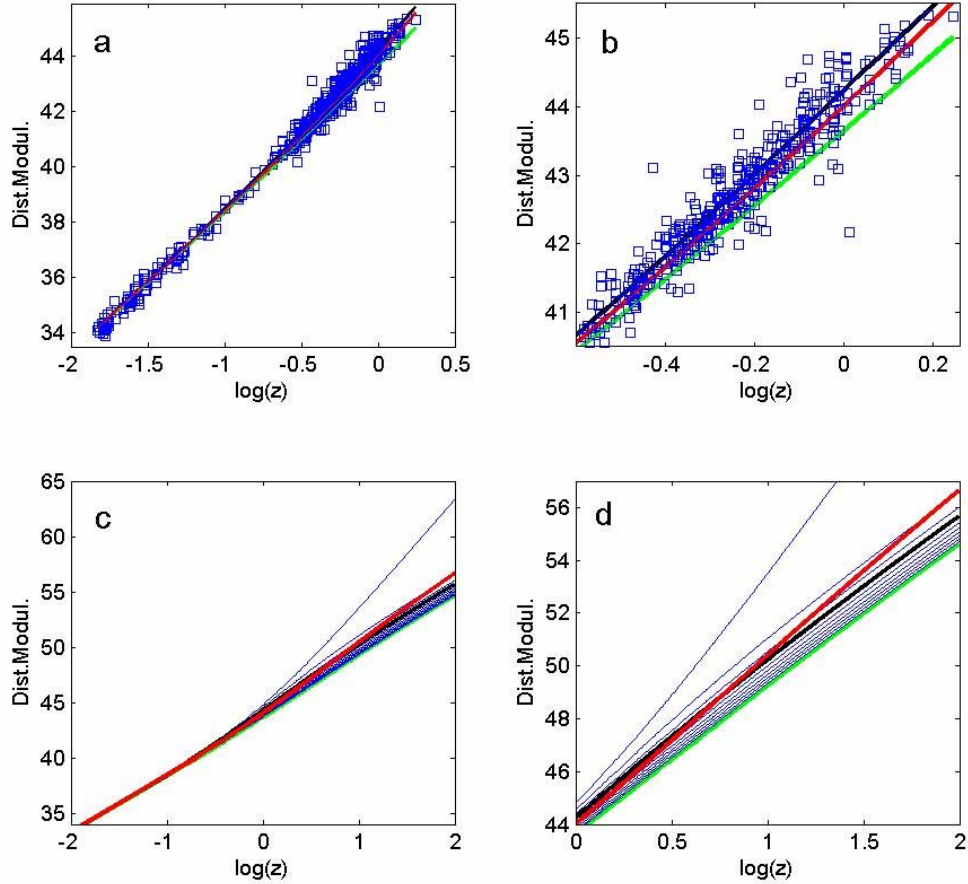

**Figure 1. Distance Modulus, μ, Versus the Log of the Redshift, z
with Data Points for 307 Ia Supernovae from the SNP Union Website**

Our fit is as good as the fit obtained by the current cosmology with its free parameters.

## 4. Dark Matter

### 4.1. In and around galaxies, space is deformed by an inhomogeneous expansion

This deformation depends on gravitational contraction, due to mass, expressed by gravitational energy density, $\epsilon_g$, and the opposing dilation, due to the dilating vibrational CMB energy density, $\epsilon_{CMB}$.

To derive the dynamic and kinematic relations that govern the motions of celestial bodies in a galaxy we consider a very simplified model. This model considers a galaxy to be a "point" mass whose formation time (the mass accretion phase) is much shorter than its present age. In other words, we assume that the galaxy was formed "instantly" at time $t_0$, when the scale factor was $a_0$, possessing its final mass value. Note that in this section $a_0$ is the scale factor value at the time of the galaxy formation and not the present value. The redshifted galactic light recorded now left the galaxy at cosmic time $t_z$, when the scale factor was $a_z$. We divide the space around a galaxy into three regions according to the relative values of $\epsilon_g$ and $\epsilon_{CMB}$:

a. **From the center of a galaxy to $R_0$, where $\epsilon_g \geq \epsilon_{CMB}$**

   $R_0$ is the distance at which, initially, at the time of formation $\epsilon_g = \epsilon_{CMB}$. In this region, the local contraction of space by the mass of the galaxy is stronger than the opposing dilation caused by the CMB. Space expansion is inhibited in this region, and hence Newtonian gravitation is applicable.

b. **From $R_0$ to R**

   R is the distance for which $\epsilon_g$ was equal to $\epsilon_{CMB}$ at the time of emission of a photon that reaches us now.
   In this region, equilibrium was first attained at a distance $R_0$, at the time, $t_0$, of formation of the galaxy. The expansion of the surrounding space beyond $R_0$, due to the expansion of the universe, lowered the $\epsilon_{CMB}$, and hence equilibrium was reached for t > $t_0$, at a greater distance $r(t) > R_0$. This is an ongoing process in which the region surrounding $R_0$ grows with time, with an ever-increasing value of the scale factor. Light that reaches us now, left the galaxy at time $t_z$. Equilibrium, $\epsilon_g = \epsilon_{CMB}$, at this time, occurred at a distance R from the center of the galaxy. Space density in the region between $R_0$ and R is <u>frozen</u>, since $\epsilon_g > \epsilon_{CMB}$. Space density at $R_0$ is larger than at R. In this region, RCs are flat, as our Extended Newtonian Gravitational Law predicts, see below. This region is the "DM Halo".

c. **From R onwards**
   In this region, where $\epsilon_g < \epsilon_{CMB}$, rotational velocities decline.

To express this inhomogeneity we introduce a scale factor $a(r,t)$ that depends not only on time, t, but also on the distance, r, from the center of a galaxy.

### 4.2. We extend the Newtonian gravitational field equation by taking into account space deformation due to its expansion

Newtonian gravitation is an approximation since, unlike GR, it does not take into account, in the calculation of the gravitational flux density, the deformation (curving) of space around a mass. However, neither theory considers the inhomogeneous strong deformation of space around a mass due to space expansion. Here we derive the flux density / gravitational field for this case.

Let d be the distance between two points at a distance r from the center of a galaxy, as measured by the fixed yardstick of an observer located at the distance $R_0$, where the scale factor is $a(R_0,t)$. For an observer with a fixed yardstick at a distance $r > R_0$ from the center, where the local scale factor is $a(r,t)$, the measured distance is d', where:

$$(20) \quad d' = d \cdot \frac{a(R_0,t)}{a(r,t)}$$

In space that is contracted gravitationally: $d' < d$ since $\frac{a(R_0,t)}{a(r,t)} < 1$.

This also holds for space that is both contracted gravitationally and is expanded inhomogeneously with r.

The area of a virtual spherical shell of radius r in deformed space, with the scale factor, a(r,t), is $A'$. The area of a spherical shell of the same radius in un-deformed space, in which the value of the scale factor, a($R_0$,t), is uniform, is A, where $A'$ is related to A as follows:

(21) $$A' = A \cdot \left[\frac{a(R_0,t)}{a(r,t)}\right]^2$$

In an expanded space, the scale factor grows from center outwards. Therefore our extended Gauss Theorem for the case of deformed space implies that the field strength, $E_g(r,t)$, which is the flux density perpendicular to the shell, is larger than the field strength for un-deformed (flat) space:

(22) $$E_g(r,t) = \frac{GM}{r^2}\left[\frac{a(r,t)}{a(R_0,t)}\right]^2$$

This is our Extended Newtonian Gravitational field equation.
By introducing the scale factor, a(r,t), equation (22) takes into account the curving of space by mass - by its direct contraction of space as expressed by GR as well as by its effect in modifying space expansion. The contribution of space expansion to the variation of the scale factor as a function of r is orders of magnitude greater than the intrinsic contribution by the mass, as expressed by GR. Therefore, in regions of inhomogeneous expanded space, equation (22) is more applicable than GR.

$R_0$ is the distance from the center of a galaxy at which $\epsilon_g = \epsilon_{CMB}$ at the time $t_0$ of its formation. Let $g_0$ be the central acceleration at this point.

Radiation density, like the CMB, which is homogeneous throughout space, including the interiors of "DM halos" (Granitt et al, 2008) is reduced with expansion. For all $t_2 > t_1$, where $a(t_2) > a(t_1)$, $\epsilon_{CMB}(t_2)$ is related to $\epsilon_{CMB}(t_1)$, as follows:

(23) $$\epsilon_{CMB}(t_2) = \epsilon_{CMB}(t_1)\left(\frac{a(t_1)}{a(t_2)}\right)^4$$

**4.3. From the general expressions for $E_g(r,t)$ and $\epsilon_{CMB}(t)$, we derive the gravitational central acceleration, in and around galaxies for the region between $R_0$ and R**

Consider a point in the second region, $R_0 < r < R$. From equation (22) for $E_g(r,t)$ we derive the gravitational energy density of contraction, $\epsilon_g(r,t)$ in this region:

(24) $$\epsilon_g(r,t) = \frac{1}{8\pi G}E_g^2(r,t) = \frac{1}{8\pi G}\left[\frac{GM}{r^2}\cdot\left[\frac{a(r,t)}{a(R_0,t_0)}\right]^2\right]^2$$

a(r,t) is the scale factor at the distance r for all times later than, t. a($R_0$,t) is the scale factor at the distance $R_0$, which was fixed at time $t_0$, and remains the same for all times later than $t_0$. Equation (23), can thus be written for t and $t_0$ as:

(25) $\epsilon_{CMB}(t) = \epsilon_{CMB}(t_0) \cdot \left[ \dfrac{a(r,t)}{a(R_0,t_0)} \right]^{-4}$

Equating equation (25) to (24) gives:

(26) $\dfrac{a(r,t)}{a(R_0,t)} = \left[ \dfrac{8\pi G \, \epsilon_{CMB}(t_0)}{G^2 M^2} \right]^{\frac{1}{8}} \cdot r^{\frac{1}{2}}$

We designate $E_g$ by $g$ and the numerator in (26) by:

(27) $g_0^2 = 8\pi G \, \epsilon_{CMB}(t_0)$

This designation is explained at the end of this section and in the following section.

We rewrite equation (26) as:

(28) $\dfrac{a(r,t)}{a(R_0,t_0)} = \left[ \dfrac{g_0^2}{G^2 M^2} \right]^{\frac{1}{8}} \cdot r^{\frac{1}{2}}$

Substituting (28) into (22) gives:

(29) $g = \dfrac{GM}{r^2} \cdot \left[ \dfrac{g_0^2}{G^2 M^2} \right]^{\frac{1}{4}} \cdot r = \left[ (GM)^2 \cdot \dfrac{g_0}{GM} \right]^{\frac{1}{2}} \cdot r^{-1} = [(GM) \cdot g_0]^{\frac{1}{2}} \cdot r^{-1}$

Thus the gravitational central acceleration in the region R to $R_0$ is:

(30) $g = \dfrac{\sqrt{g_0 GM}}{r}$     which resembles the Milgrom (1983) relation, but is in no way related to the MOND paradigm.

Squaring equation (30), gives $g^2 / g_0 = GM/r^2$

Since $g = \dfrac{v^2}{r}$ we get: $\dfrac{v^4}{r^2} = g_0 \dfrac{GM}{r^2}$     or:

(31) $v^4 = (g_0 G) M = B \cdot M$     which is the Tully-Fisher relation.

The circular rotation velocity in this region is:

(32) $v = (g_0 GM)^{\frac{1}{4}}$

and thus <u>RC in this region is flat</u>. Section 4.8 shows that a more realistic model that takes into account the evolution of galaxies yields RCs that fit observed RCs.

From equation (27) that defines $g_0$:

(33) $\dfrac{1}{8\pi G} g_0^2 = \epsilon_{CMB}(t_0)$

Thus, $g_0$ is the field strength (central acceleration) at $R_0$, at the time, $t_0$, of formation.

Note that the region, $R_0$ to R, in which space density is frozen, grows with time. At $R_0$ space density is high – small $a(r,t)$ – and is reduced towards R – higher $a(r,t)$. At distances r > R, where $\epsilon_{CMB} > \epsilon_g$, space expands. In this region space expands homogeneously, hence the central acceleration is proportional to $r^{-2}$.

### 4.4. Some numerical results for $g_0$, B and $R_0$

- **For $g_0$**
  For galaxies formed at z ~ 3 the corresponding scale factor, a, is 0.25, (Baugh et al, 1998) and hence the time of formation, $t_0$, is ~11 BY. To obtain the value for $\epsilon_{CMB}(t_0)$ we use equation (23).
  The present value, $\epsilon_{CMB}(Now) = 4.17 \cdot 10^{-13}$ erg cm$^{-3}$, gives for $t_0$:
  $\epsilon_{CMB}(t_0) \sim 1.0 \cdot 10^{-10}$ erg cm$^{-3}$. Hence, from equation (27) we get:
  
  (34)  $g_0 = \sqrt{8\pi G \epsilon_{CMB}(t_0)} \sim 1.3 \cdot 10^{-8}$ cm s$^{-2}$.
  
  This result is close to the Milgrom (1983) "universal constant" but is not a constant at all (B. Famaey et al, 2007). Observations show that the central acceleration $g_0$ takes a wide range of values, as our expression for this parameter predicts, see Begeman et al (1991) and Scott et al (2001).

- **For B, the Tully-Fisher Parameter**
  In a large sample of galaxies, covering a wide dynamic range, McGaugh et al (2000) found the Tully-Fisher relation $M = AV^4$ between the baryonic mass, M, of galaxies and the rotation velocities at the flat edge of their RCs. The coefficient A in this relation is found empirically to be $A = 35 h_{75}^{-1} M(Sun) s^{-4}$ km$^{-4}$. Here $h_{75}$ is the Hubble constant in units of 75 (km/s$^{-1}$)/Mpc. With h = 1, A = 7 x $10^{14}$ gr s$^{-4}$ cm$^{-4}$. We write the Tully-Fisher relation, equation (32), as $V^4 = BM$, where according to equation (34) $B = G\sqrt{8\pi G \epsilon_{CMB}(t_0)}$, for $t_0$, the time of the galaxy formation. Clearly B = 1/A.
  
  Using equation (23) $\epsilon_{CMB}(t_0) = a^{-4}(t_0)\epsilon_{CMB}(Now) = (1+z_0)^4 \epsilon_{CMB}(Now)$ we obtain a fit of our B to the value 1/A for $z_0 = 4$. Baugh et al (1998) derived the value z = 3 to 3.5, for the redshift at the time of galaxy formation, from an entirely independent set of observations. Our formation time differs from that of Baugh, which is the epoch at which a galaxy first becomes detectable in optical and IR light. In this paper we define the formation time of galaxies differently, as explained in Section 4.1. It is likely that this time precedes the birth of light emitting objects in a proto-galaxy. This may explain the difference between the two results.

- **For $R_0$**
  For a given M the distance $R_0$ is:
  
  (35)  $R_0 = \sqrt{\dfrac{GM}{g_0}}$
  
  In our simplified model, this is the distance from the center of a galaxy at which the "DM halo" starts. As an example, for a galaxy formed ~11 BY ago with a bulge mass M ~ $1.3 \cdot 10^{10}$ M$_\odot$, our calculation gives $R_0$ ~ 3 KPC. Assuming similar initial conditions for the Milky Way galaxy, the above calculation of $R_0$ is supported by observations, (Gerhard, 2002). From this distance onwards, the rotational velocity increases, reaches a "plateau" and then decreases, as is indicated by observations of dispersion velocities (Battagalia, 2005).

To summarize, in our simplified model of a galaxy with mass M, the gravitational field, $E_g = g$, around a galaxy, for the three regions of an RC, is:

(36) $\quad r \leq R_0 \qquad g = \dfrac{GM}{r^2}$

(37) $\quad R_0 < r \leq R \qquad g = \dfrac{\sqrt{g_0 GM}}{r}$

In reality, g in the first and second regions depends on the mass distribution and the history of the galaxy formation.

(38) $\quad r > R \qquad g = \dfrac{GM}{r^2} \left[ \dfrac{a_z}{a_0} \right]^2$

$a_z$ and $a_0$ are defined in Section 3.

## 4.5. The gravitational potential is modified by space expansion

By integrating equation (30) for g, we get the potential difference.

(39) $\quad \varphi(r) - \varphi(R_0) = \int_{R_0}^{r} g\, dr = \sqrt{g_0 (GM)} \cdot \int_{R_0}^{r} r^{-1} dr = \sqrt{g_0 (GM)} \cdot \ln \dfrac{r}{R_0}$

This potential difference is only valid in the region between $R_0$ and R. Since:
$\quad \varphi(R_0) = -GM/R_0 \quad$ (in reality, $\varphi(R_0)$ depends on the mass distribution) we get:

(40) $\quad \varphi(r) = \sqrt{g_0 (GM)} \cdot \ln \dfrac{r}{R_0} - \dfrac{GM}{R_0}$

Figure 2 illustrates a typical potential in the three regions above as a function of the distance from the center of the galaxy and for different times.

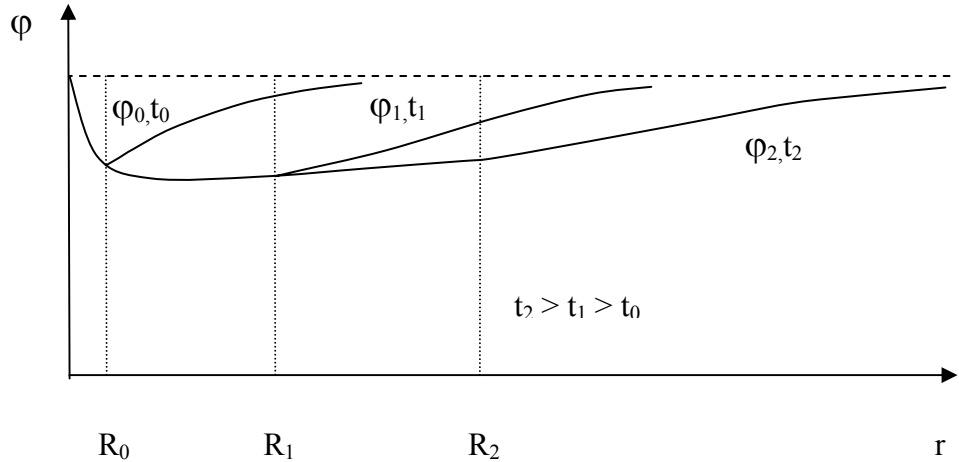

**Figure 2. The Gravitational Potential of a Galaxy in an Expanding Universe**

The curves for φ in Figure 2 show that, with time, the zone of flat RC grows. This means that with time "DM halos" should grow. Observations confirm this result (Massey et al, 2007).

### 4.6. The gravitational potential in an expanding universe explains the enhanced gravitation lensing

A point mass, M, which serves as a lens, deflects a light beam with an impact parameter, b, at the following deflection angle:

$$(41) \quad \alpha = \frac{4GM}{c^2 b} = \frac{4}{c^2} \varphi$$

where φ is the gravitational potential at a distance b from M (Carroll, 2004, Sec. 7.3 and Sec. 8.6).

However, the potential in the zone of flat RCs around M, expressed by equation (40), yields, for large impact parameters, a much larger deflection of light beams.

### 4.7. "DM Halos" are zones of condensed space

DM halos can be detached from fast moving galaxies like the "bullet cluster" 1E0657-56, (Clow et al 2004). We are thus lead to the conclusion that the two ways by which mass deforms space differ from each other as follows:

- **Elastic deformation by the presence of mass alone**
  GR states that space deformation is gravity, i.e., in the vicinity of masses, space cells are contracted. This contraction is elastic - remove the mass and space resumes its original geometry.

- **Non-elastic deformation due to space expansion around a mass**
  In addition to the above elastic deformation, space is also deformed by the inhomogeneous space expansion around the mass, caused by the mass. Such deformation is observed as a DM halo, as shown above. However, in contrast to elastic deformation, the halo does not follow a moving mass and retains its geometry. A DM halo, without the presence of a mass, is subject to Hubble expansion. Elastic deformation is orders of magnitude smaller than the non-elastic deformation (Section 2.3) accumulated over cosmological time.

### 4.8 Galactic Evolution is taken into account in a more realistic model that yields realistic RCs

Galaxies attain their observed baryonic masses and mass distribution during a time span that is a fraction of the age of the universe (Searle and Zinn 1978, Martinez-Delgao et al 2008). The evolution of a deformed space halo around a galaxy is determined by both the decline of $\in_{CMB}$ with space expansion, and the history of the accumulation or loss of mass by the galaxy.

To account for this history we introduce the following functions and parameters:

- $\mu(t) = M(t)/M_0$ describes the mass evolution of the galaxy, normalized to the galactic mass at formation.

- $\chi(r) = M(T,r)/M_0$ represents the mass distribution in a finalized galaxy, as observed. Specifically, $\chi(r)$ is the (normalized) mass of a sphere of radius $r$ in the observed galaxy. For $r > R_T$, where $R_T$ is the radius of the spherical distribution of the baryonic matter of the mature galaxy, $\chi(r) = \mu_T$. Here $\mu_T = M_T/M_0$ is the observed (normalized) mass of the galaxy.

- $\xi = R_T/R_0$ is the ratio of $R_T$, the radius of the mature galaxy, to $R_0$, the radius of the infant galaxy.

- $\rho = r/R_T$ expresses distances from the center of a galaxy with a dimensionless normalized radial coordinate

Our simplifying assumptions are:
- After the formation of the galaxy, at time $t_0$, all accreted matter is distributed instantly according to the observed final distribution.
- The radius of the frozen sphere of space is always larger than the instantaneous radius of the galaxy.

This model, using equations (28) and (35), and designating $a(R_0, t_0) = a_0$, gives:

$$(42) \quad \frac{r(t)}{R_0} = \mu^{\frac{1}{2}}(t) \left[\frac{a(t)}{a_0}\right]^2.$$

This can also be written for a piecewise linear $\mu(t)$ as a function of a:

$$(43) \quad \mu^{\frac{1}{2}}(a)\left(\frac{a}{a_0}\right)^2 - \rho = 0$$

We now compute the gravitational field strength at every radius $\rho$, $\rho_0 \leq \rho \leq \rho_z$ as follows:

For each $\rho$ we consider expression (43) as an equation for $a = a(\rho)$, recalling that $a_0 = (1+z_0)^{-1}$. Its solution is the cosmic scale factor, $a$, at which the radius of the frozen sphere arrives at the distance $\rho$ from the center. From that moment on, space expansion is frozen at this point with this value $a$ as the local scale factor.

Substituting this value, $a(\rho)$, in equation (22), where $E_g = g$, and from the equation for the circular rotation velocity, $v = \sqrt{rg}$, we obtain:

$$(44) \quad v(\rho) = \sqrt{\frac{GM_0}{R_0}} \sqrt{\frac{\chi(\rho)}{\rho}} \left(\frac{a}{a_0}\right).$$

At the outskirts of the galaxy, for $\rho > \rho_z$ where $\rho_z$ is the radius of the frozen sphere at the time the recorded photon left the galaxy, the Extended Newtonian expression is:

$$(45) \quad v(\rho) = \sqrt{\frac{GM_0}{R_0}} \sqrt{\frac{\chi(\rho)}{\rho}} \left( \frac{a_z}{a_0} \right).$$

Here, as above, $a_z = (1+z)^{-1}$ is the value of the cosmic scale factor for the measured redshift of the galaxy. The Newtonian expression for the rotation velocity, for all $\rho \geq \rho_0$ is:

$$(46) \quad v_N(\rho) = \sqrt{\frac{GM_0}{R_0}} \sqrt{\frac{\chi(\rho)}{\rho}}$$

### 4.9 We compare our theoretical RCs with observed RCs

We present three theoretical RCs, obtained from our model with no need of any mass in addition to that of the observed luminous baryonic matter and compare them qualitatively with observed RCs of three galaxies.

In our simple model, an initial accretion phase is followed by one or two phases of mass accretion or loss (due, for example, to SN explosions or stellar winds). In this model, mass accretion or loss occurs at a constant rate. The duration of each phase, and the accretion or loss rates, are free parameters of the model.

The large frames on the left-hand side of Figure 3 present the observed RCs of the galaxies NGC 2903, NGC 3657 and UGC 4458. (de Blok et al 2008, Milgrom 2008, and Sanders and Noordermeer 2007, respectively).

The thick lines in the curves on the right-hand side are the theoretical model curves. The thin lines are the corresponding Newtonian curves. The x-axis is the normalized radial distance $\rho$, as defined above. The numbers on the y-axis are dimensionless, expressing rotation velocities in units of $\sqrt{GM_0/R_0}$.

The curves below each RC plot show the assumed evolution of the models used to generate the theoretical curves. They show for each model the (normalized) mass of the galaxy as a function of cosmic time t (left curve), taking as unity the present age of the universe, T = 1, and redshift z (right curve).

The theoretical RCs fit the observed RCs.

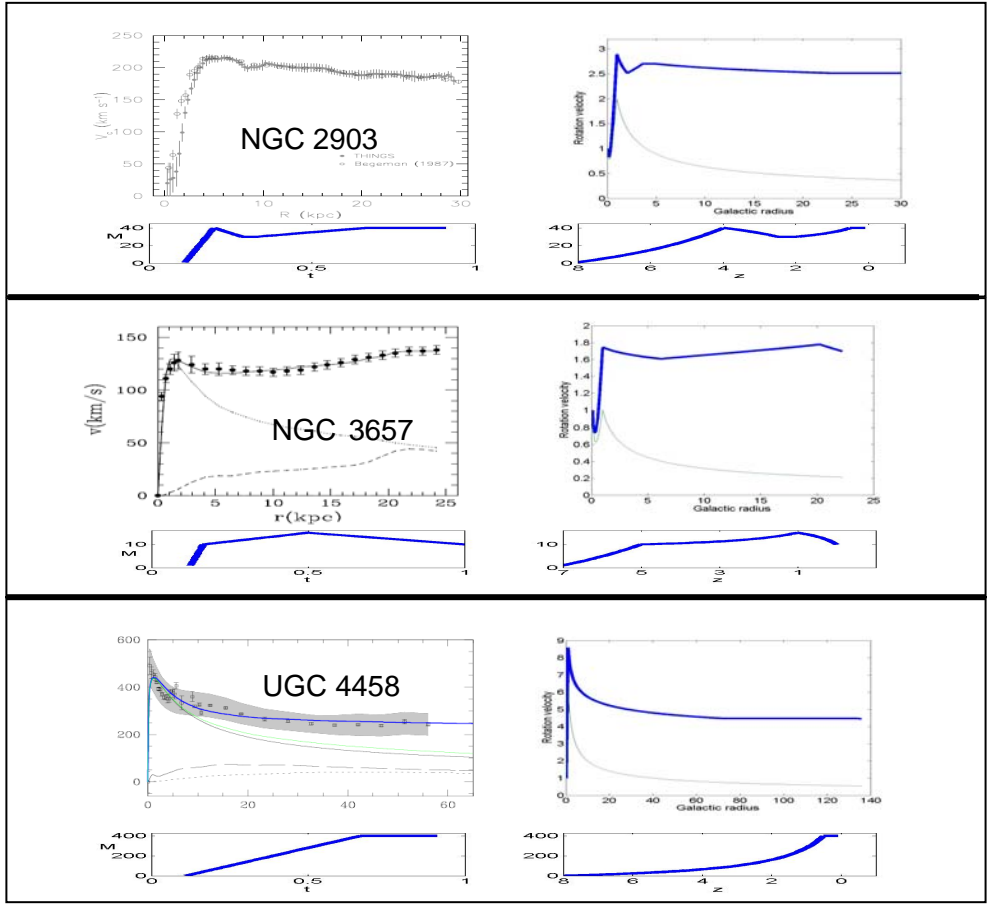

**Figure 3. Comparison of our theoretical RC with observed RCs**

The similarity in the profile of the theoretical plots to that of the observed RCs is evident. However, we do not claim that the parameters of our model necessarily characterize the true histories of the three galaxies. They are not even determined uniquely by the profiles of the curves alone. As discussed above, our model is over simplistic and does not take into account observed data of the real galaxies such as their surface brightness. The purpose of this exposition is merely to demonstrate qualitatively that our theory is capable of explaining observed RCs of galaxies, even at very large distances from their centers, with no need of any mass in addition to that of the observed luminous baryonic matter. To establish the fit on a more quantitative footing, much more work is required to develop equations of non-spherical mass distributions, which should also incorporate data of measurements in real galaxies.

## 5. Summary

We show that in the inter-galactic space the global gravitational energy density, $\epsilon_g$, equals the Cosmological Microwave Background (CMB) energy density, $\epsilon_{CMB}$, and that both depend on $a^{-4}$. This equality, $\epsilon_g \approx \epsilon_{CMB}$, implies that $\dot{a} = \text{const}$, ($\ddot{a} = 0$).

The above leads to H(z) = $H_0$(1+z) and hence to the Distance Modulus $d_L \propto \ln(1 + z)$. This result is supported by its fit, without any free parameters, to data from observations of hundreds of Ia supernovae. This validates our theoretical result that $\dot{a} = \text{const}$.

We consider the dilation (negative curving) of space by the $\epsilon_{CMB}$ and the contraction (positive curving) of space as expressed by $\epsilon_g$. The interplay between $\epsilon_{CMB}$ and $\epsilon_g$ in, and around, galaxies explains the local inhomogeneous space expansion.

We extend Newton's field equation to account for space deformation caused by this local inhomogeneous expansion. This leads to a theoretical derivation of the gravitational central acceleration in, and around galaxies, and to the Tully-Fisher kinematic relation.

Our theoretical results fit observations and thus explain the flattening of Rotational Curves.

We have shown that the interplay between $\epsilon_{CMB}$ and $\epsilon_g$ is sufficient to account for cosmological and astrophysical phenomena currently attributed to Dark Energy and Dark Matter, and have thus dispelled their mysteries.

## Acknowledgments

We would like to thank Roger M. Kaye for his linguistic contribution and technical assistance.